\documentclass[
twocolumn,
longbibliography,
showpacs,                     %  Add 'showpacs' option to make PACS codes appear
showkeys,                  %  Add 'showkeys' option to make keywords appear
secnumarabic,
amssymb,
nobibnotes,
floatfix,
aps,
pre]{revtex4-1}
\usepackage{graphicx}
\usepackage{placeins}
\usepackage{latexsym}
\usepackage{amsmath}
\usepackage{verbatim}

\begin{document}

\preprint{Inverse Ising}

\title{Detecting Multi-Spin Interactions\\ in the   Inverse Ising Problem}

\author{Joseph Albert}
\email{ joseph@psu.edu}
% \email{jcalbert@alumni.cmu.edu}
\affiliation{
 Physics Department,
 Carnegie Mellon University,
 Pittsburgh, Pennsylvania 15213, USA 
}
 
\author{Robert H. Swendsen}
 \email{swendsen@cmu.edu}
\affiliation{
 Physics Department,
 Carnegie Mellon University,
 Pittsburgh, Pennsylvania 15213, USA 
}

\date{\today}

\begin{abstract}
While the usual goal 
in Monte Carlo (MC) simulations of Ising models
is the efficient generation of spin configurations
with  Boltzmann probabilities,
the inverse  problem is to 
determine the coupling constants 
from a given set of 
spin configurations.
Most recent work has been limited to 
local magnetic fields and pair-wise interactions.
We have extended solutions to multi-spin interactions,
using correlation function matching (CFM).
A more serious limitation of previous work 
has been the uncertainty 
of whether a chosen set of interactions 
is capable of faithfully representing real data.
We show how 
our confirmation testing method 
uses an additional MC simulation
to  detect significant interactions 
that might be missing in the assumed
representation of the data.
\end{abstract}


\maketitle

\section{Introduction}

In many  fields,
 such as biology, sociology,
and neuroscience,
obtaining information about 
underlying interactions
between components of a system from observed correlations 
can clarify the structure of the 
system\cite{Schneidman,Lezon_2006,Cocco_2009,Tkacik_Schneidman_Berry_Bialek_2006,Tkacik_Schneidman_Berry_Bialek_2009,Weigt_2008,Schug_2009,Morcos_2011,Aurell_Ekeberg}. 
This reconstruction of cause from consequence
is known as an inverse problem.
Because of its relative simplicity, 
the inverse Ising model 
has become
a standard  test case 
for the development of methods to deal with  
intrinsically complex inverse problems.
In 1984,
a numerical solution to the problem was found 
by  correlation function matching
(CFM), 
using an identity due to Callen\cite{RHS_MCRG_Ks,RHS_MCRG_Ks_d=2,RHS_MCRG_Ks_d=3,Callen}.
At the time, the solution was only applied to transitionally invariant
problems,
but as we show below,
the modifications 
to remove  this restriction 
are trivial.

Recently,
equations  
originally found 
with CFM  
were rediscovered
by Aurell and Ekeberg,
starting from  
  different  principles
  (pseudo-likelihood),
and successfully applied  to 
the Sherrington-Kirkpatrick (SK) model\cite{Aurell_Ekeberg}.
Their approach
has the advantage
of exhibiting the relationship of the solution 
to a Bayesian probability distribution 
on the space of the coupling constants.
The CFM approach,
on the other hand,
 clarifies the relationship between 
extracting information from the 
configurations and 
making inferences about the  original  coupling constants.

Recent work 
on inverse problems  
has been largely  restricted to pairwise interactions,
as in the SK model.
We have extended it to include multi-spin interactions\cite{RHS_MCRG_Ks,RHS_MCRG_Ks_d=2,RHS_MCRG_Ks_d=3,Callen},
but
there is still a
  question 
of whether 
a given set of real data 
can be faithfully represented by
the chosen set of interactions.
We answer this question 
by introducing a 
confirmation phase into our computations,
using a new Monte Carlo (MC) simulation
with the fitted coupling constants.
By examining
 correlation functions 
\emph{that were not used in the inverse solution},
we show  that 
differences
between 
 the new MC simulation 
and the original data
 reveal 
 neglected interactions.
This ``confirmation testing''
provides  a straightforward 
way of determining whether
more  interactions are 
 needed for a faithful representation
 of the data, 
without 
having to perform
 a full computation 
 of the coupling constants  
 for the additional interactions.
We will next describe the CFM equations that provide the basis for confirmation testing.

\section{CFM  equations}

To express a general, multi-spin  Ising interaction,
let $\alpha$ be a subset of 
$m_{\alpha}$ spins,
and
define  the product of all  spins in 
$\alpha$ as 
\begin{equation}\label{S alpha}
S_{\alpha}
=
\prod_{j=1}^{m_{\alpha}} \sigma_{j}    .
\end{equation}
For each operator 
$S_{\alpha}$,
we will assign
 a corresponding dimensionless  coupling constant,  
$K_{\alpha} = \beta J_{\alpha}$,
where
$k_B$ is Boltzmann's constant,
  $T$ is the temperature,
  and
$\beta = 1/ k_B T$.
The corresponding 
term in the dimensionless Hamiltonian 
$\mathcal{H}  = - \beta H$,
(where $H$ is the usual Hamiltonian)
associated with  
$\alpha$
is
$K_{\alpha} S_{\alpha}$.
The full  dimensionless Hamiltonian can then be written as 
a sum over the set of all spin products, 
\begin{equation}\label{H = sum S alpha}
\mathcal{H} 
=
\sum_{\alpha}
K_{\alpha}
S_{\alpha} ,
\end{equation}
where 
the set 
$\textbf{K}=\{K_{\alpha}\}$,  
are the  true values of the coupling constants.

We define  an operator,
$\mathcal{H} _{\ell}$,
 that includes  all 
terms in the Hamiltonian  containing 
a specific spin, 
$\sigma_{\ell}$\cite{RHS_MCRG_Ks,RHS_MCRG_Ks_d=2,RHS_MCRG_Ks_d=3}.     
If 
$\sigma_{\ell} \in \alpha$,
we also define an  operator 
$\widehat{S}_{\alpha, \ell}$
that omits the spin $\sigma_{\ell}$.
\begin{equation}\label{S alpha ell}
\widehat{S}_{\alpha, \ell}
=
S_{\alpha} /  \sigma_{\ell}
\end{equation}
If $\sigma_{\ell} \notin \alpha$,
$\widehat{S}_{\alpha, \ell}=0$.
 The sum of all terms     in the  Hamiltonian
  that contain the 
  ``central''
  spin
$\sigma_{\ell}$
is then
\begin{equation}\label{H ell}
\mathcal{H} _{\ell}
=
\sigma_{\ell}
\sum_{\alpha}
K_{\alpha}
\widehat{S}_{\alpha,\ell}     .
\end{equation}

The CFM method is based on  fitting   the   correlation functions
$c_{\alpha} \equiv  \left\langle   S_{\alpha}  \right \rangle$
using an identity
due to      Callen\cite{Callen}.  
\begin{equation}\label{Callen identity 1}
c_{\alpha}     \left( \textbf{K} \right)
\equiv
\left\langle     \widehat{S}_{\alpha,\ell}  
\tanh \left[  \sum_{\gamma}
K_{\gamma}
\widehat{S}_{\gamma,\ell}   \right]
 \right\rangle
 =
 \left\langle   S_{\alpha}  \right \rangle
\end{equation}
For comparison with earlier work,
Eq.~(\ref{Callen identity 1}) 
corresponds to 
Eq.~(14)
in Ref.~\cite{RHS_MCRG_Ks}.
Aurell and Ekeberg
considered 
$P(\sigma_j) = (\sigma_j +1)/2$
instead of 
$\sigma_j $.
%as in   Eq.~(\ref{Callen identity 1}).
Since
the probability that a spin is positive is 
\begin{equation}
P(+1)
=
 \left(1 - \exp \left[ - 2  F_{\ell}  \right] \right)^{-1} 
=
\left( \tanh \left[  F_{\ell}  \right] + 1 \right)/2 ,
\end{equation}
where 
$F_{\ell}  = \sum_{\gamma}
K_{\gamma}
\widehat{S}_{\gamma,\ell} $,
Eq.~(\ref{Callen identity 1}) 
is equivalent to  
Eq.~(4)
in Ref.~\cite{Aurell_Ekeberg}.

Eq.~(\ref{Callen identity 1})
is exact
for the correct values of the coupling constants
and the
exact correlation functions.
Unfortunately, we never have 
the exact values of the correlation
functions 
$c_{\alpha} =\left\langle   S_{\alpha}  \right \rangle$.
Data  
always come   from a finite sample,
which we will take to be  
 $N_{\textrm{MC}}$
spin
configurations.
We  denote the corresponding 
approximate correlations functions as 
$\widehat{c}_{\alpha}=\left\langle   S_{\alpha}  \right \rangle_{MC}$.
We can then use a modification of
Eq.~(\ref{Callen identity 1})
to find  a  set of  approximate coupling parameters
$\widetilde{\textbf{K}} = \{ \widetilde{K}_{\gamma} \}$ 
to fit the MC correlation functions.
\begin{equation}\label{Callen identity 2}
\widehat{c}_{\alpha}   \left( \widetilde{\textbf{K}}  \right)
\equiv
\left\langle     \widehat{S}_{\alpha,\ell}  
\tanh \left[  \sum_{\gamma}
\widetilde{K}_{\gamma}
\widehat{S}_{\gamma,\ell}   \right]
 \right\rangle_{MC}
 =
 \left\langle   S_{\alpha}  \right \rangle_{MC}
\end{equation}
The equality in 
Eq.~(\ref{Callen identity 2})
will only hold 
for specific values of  the coupling parameters
$\widetilde{\textbf{K}} $.
If  trial values for the set 
$\widetilde{\textbf{K}} $
differ from the best-fitting value by 
$\delta \widetilde{\textbf{K}}= \{ \delta K_{\alpha} \} $,
an improved estimate can be obtained 
from a linearized approximation
for the deviations from the best-fitting values.
\begin{equation}\label{Callen identity 3}
\widehat{c}_{\alpha}   \left( 
\widetilde{\textbf{K}} 
+\delta \widetilde{\textbf{K}} 
\right)
 -
 \left\langle   S_{\alpha}  \right \rangle_{MC}
\approx
\sum_{\beta}
\frac{ \partial  \,  \widehat{c}_{\alpha}   \left( \widetilde{\textbf{K}}  \right)
}{
\partial K_{\beta}
}
\,
\delta  \widetilde{K}_{\beta}
\end{equation}
%
%The correlation function differences on the left side of Eq.~(\ref{Callen identity 3})  will be the basis for confirmation testing.
The derivatives in 
Eq.~(\ref{Callen identity 3})
are given by 
\begin{equation}\label{derivative 1}
\frac{ \partial  \, \widehat{c}_{\alpha}   \left( \widetilde{\textbf{K}}  \right)
}{
\partial K_{\beta}
}
=
\left\langle     \widehat{S}_{\alpha,\ell}  
    \widehat{S}_{\beta,\ell}  
    \,
\textrm{sech}^2 \left[  \sum_{\gamma}
\widetilde{K}_{\gamma}
\widehat{S}_{\gamma,\ell}   \right]
 \right\rangle_{MC}
\end{equation}
Eq.~(\ref{Callen identity 3})
is  iterated until 
convergence,
which is quadratic in the absence of a degeneracy.
Again for comparison with earlier work, 
Eqs.~(\ref{derivative 1})
corresponds to 
Eq.~(15)
in Ref.~\cite{RHS_MCRG_Ks},
and 
Eq.~(7)
in Ref.~\cite{Aurell_Ekeberg}.
Note that for an interaction $\alpha$ with 
$m_{\alpha}$ 
spins,
this procedure 
produces $m_{\alpha}$ 
 values of $\widetilde{K}_{\alpha}$,
 one for each choice of the ``central'' spin 
 $\sigma_{\ell}$.
Before going on to confirmation testing, 
we next describe the application 
of the CFM solution to the inverse Ising problem, 
along with limitations that not only CFM, but any inverse method, 
will have with respect to recovering the true coupling constants. 
A virtue of the CFM is that it exposes these limitations clearly.

%%%%%%

\section{The limited information contained in a set of configurations}\label{section: info contained}

There is an important distinction  between 
extracting  information contained in the configurations
and inferring the values of the true couplings.
For example,
if
$\mathcal{H} = h \sigma$,
$\langle \sigma \rangle_{\text{exact}} 
=
\tanh ( h_{\text{exact}})$  
is
the 
 exact average value of a spin $\sigma$
 in a dimensionless   field $h_{\text{exact}}$.
Given  
$\langle \sigma \rangle_{\text{MC}}$  
from an MC simulation,
it is easy to find
 an effective magnetic field,
$h_{\text{eff}} =  \tanh^{-1} \langle \sigma \rangle_{\text{MC}}$,
that reproduces it 
 to arbitrary accuracy.
Since 
 $\langle \sigma \rangle_{\text{MC}}$
 is not exactly equal to 
  $\langle \sigma \rangle_{\text{exact}}$,
  $h_{\text{eff}}$ will differ from 
  $h_{\text{exact}}$.
The 
  uncertainty in inferring $h_{\text{exact}}$  
is given by 
$\delta h
\approx
 \cosh ( h )  / \sqrt{N_{MC}}$.
For small values of $h$,
the error  
 is approximately equal to
  the 
minimum error,
   $  \delta h \approx \delta h_{\text{min}} = 1/  \sqrt{N_{MC}}$.
Our simulation results have shown that the errors 
in estimating coupling constants at high temperatures	
are very close to the minimum error,
even for large numbers of spins.

This simple estimate of
$h_{\text{eff}} $
breaks down for 
$ \langle \sigma \rangle_{\text{MC}}   = 1$
because it 
would imply that  $h_{\text{eff}}   = \infty$.
 A simple  Bayesian argument
suggests
 replacing 
$\langle \sigma \rangle_{\text{MC}}$ 
by 
$1-2/N_{\text{MC}}$,
which gives a finite value for 
$h_{\text{eff}} $.
Although this is a  coarse 
method,
it is quite effective for improving results.
Aurell and Ekeberg
used a similar strategy,
but took 
the factor to be 
$0.999$ for all values  
of $N_{\text{MC}}$\cite{Aurell_Ekeberg}.

The limited information contained in the configurations is illustrated by 
  the Sherrington-Kirkpatrick (SK) model of a spin glass\cite{SK}.
In this model,
there are $N$ spins,
$\sigma =
\{
\sigma_j = \pm 1  \vert j=1,2,\dots,N  \} $,
and
the Hamiltonian is  
\begin{align}\label{H random Ising 1}
H(\sigma) = & 
 - \sum_j  b_j  \sigma_j 
- \sum_{j>k} J_{j,k}  \sigma_j  \sigma_k  ,
\end{align}
where the  couplings 
$ J_{j,k}  $
have  a quenched Gaussian distribution of width 
$J/\sqrt{N}$.
The local magnetic 
fields
$  b_j $
can either be set equal to zero
or given independent  quenched values.
The corresponding  dimensionless coupling constants 
are
$K_{j,k}= \beta J_{j,k}$
and
$h_j = \beta b_j$.

The SK model is known to have 
a  rugged energy landscape
and
a spin-glass 
phase transition at $k_B T_c=J$.
This makes it very difficult to 
generate independent configurations 
 at  low temperatures,
which limits the  information 
carried by the configurations.
However, it does not affect our ability 
to extract whatever information there is.
While we must be careful in interpreting 
our results,
predictions cannot be made 
  more accurate 
without additional information.

For large systems
at high temperatures,
a different limitation comes from the minimum error
for the correlation functions.
Since the magnitude of the couplings 
goes as $1/\sqrt{N}$,
the maximum temperature
for which it is possible to determine 
the couplings 
to an accuracy of $\epsilon = \delta K_{\ell,j}/K_{\ell,j} $
is
\begin{equation}
 T_{max} 
\lessapprox
\frac{ \epsilon J}{k_B} 
 \sqrt{\frac{N_{MC}}{N}}  .
\end{equation}

To demonstrate that  large coupling constants 
are not intrinsically difficult to determine
-- except for the factor of $\cosh(h)$ in the errors --
we've carried out simulations with optimal sampling,
that is,
  choosing independent random  values for the spins 
at  neighboring sites of 
a central site $\ell$.
The values of  $\sigma_{\ell}$
are then chosen with the thermal probability
for a trial set of $K_{\ell,j}$'s.
While the errors increase  at low temperatures,
 good estimates of the 
original couplings can still be obtained 
for an SK distribution of quenched couplings 
down to $T=0.05 \, T_c$,
as shown in Table \ref{toy table 1}.

\begin{table}[h]
\caption{ Errors in estimates of original couplings from
MC simulations for the
toy model with random sampling.
  $n=N-1 =49$ neighboring spins 
were independently 
 assigned the values $\pm 1$ 
with equal probability.
The true couplings were generated randomly
from a Gaussian distribution
with the width $ \beta J / \sqrt{N} $ .
The width of the actual distribution is given as $\Delta K$
in the second column.
For these simulations,
$N_{\text{MC}}=10^5$.
The RMS error in the estimated couplings,
given in the third column as 
$\delta K_{\ell,j}$, 
was found from the results of  ten independent  trials.
The errors in the $K$'s were also compared to the minimum error
$\delta_{min}=1/\sqrt{N_{\text{MC}}}=0.0032$
in the last column.
}
\begin{center}
\begin{tabular}{|c|c|c|c|c|c|c|c|}
\hline
T &

$\Delta K$
 &

$ \delta K_{\ell,j}  $     &

$ \delta K_{\ell,j}  / K$   

  &

$ \delta K_{\ell,j}  / \delta_{min}$        \\
  
 \hline
 10.0	&
    0.0163 &
0.0099   &	  0.609  & 0.99	\\
 
  2.0	& 
    0.0685 &
0.0034   &	  0.50  & 1.08	\\

  1.0	& 
    0.1397 &
0.0041   &	  0.029  & 1.29	\\

  0.5	&
    0.2850 &
0.0053   &	  0.0186  & 1.67	\\

  0.2	& 
   0.7186 &
0.0095   &	  0.0132  & 3.01	\\

  0.1	&
1.3464 &
0.0160   &	  0.0119  & 5.05	\\

  0.05  &
     3.3452 &
0.0660   &	  0.0197  & 20.9	\\

 \hline
 \end{tabular}
\end{center}
\label{toy table 1}
\end{table}%

The difficulties in extracting information from configurations 
generated 
below the critical temperature 
of the SK model 
are not due to any defect in the method of solution;
the information is simply not available.

Because 
the low-temperature SK model has
 extremely long correlation times, 
MC simulations will usually only sample 
near  a local free-energy minimum.
For those states,
it is common for some of the correlation functions
 to lock into values of 
$\pm1$.
When this happens,
little information can be obtained 
about the corresponding interaction.

If we have the option to change the 
sampling method,
 restarting the simulations
with different random initial conditions
will generally improve results.
Even though it is not possible 
to generate a complete sampling of a low-temperature spin glass
in this manner,
it can improve estimates of coupling constants.

The twin features of having small coupling constants for 
large systems and high temperatures,
and a phase transition that 
limits 
information 
at low temperatures,
leave only a small range of parameters 
for testing.  
This led us to use 
short-range  models  
with multi-spin  interactions
to illustrate confirmation testing,
as discussed in the following section.

\section{Confirmation Testing }\label{section: conf test}

Any solution of the inverse Ising problem assumes a certain set of interactions
that might be non-zero.  
When using confirmation testing, after fitting the coupling constants for those interactions, 
we perform an additional simulation using 
the fitted coupling  constants to generate a new set of configurations.  
We then compare the correlation functions in the new set of configurations 
with those in the old set.  
If they match within the statistical errors discussed above, 
we have confirmed our assumptions.  Deviations may reveal important interactions that were initially missing.

To illustrate confirmation testing, 
we have done MC simulations of a Hamiltonian with 
 magnetic fields,
nearest-neighbor pair interactions,
and   four-spin interactions
 on  nearest-neighbor plaquettes
 on a $32 \times 32$
  lattice,
 with $N_{MC}=10^5$.

\begin{figure}[ht]
\begin{center}
 \includegraphics[width=\columnwidth]{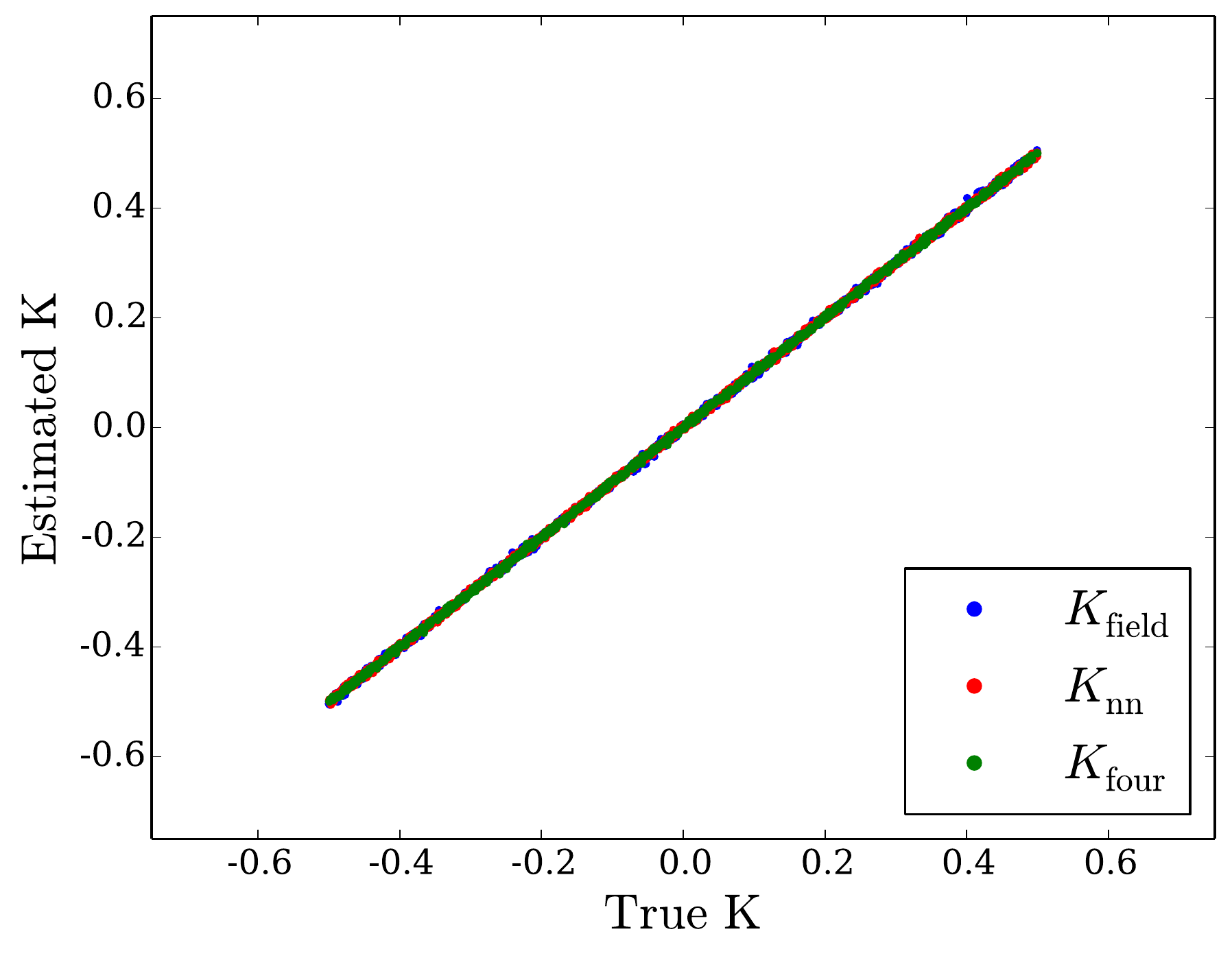}
  \caption{
  Plot of (well) estimated values of 
  $h_{field}$, $K_{nn}$, and $K_{four}$
  vs. their true values
  using data from an  MC simulation
on a $32 \times 32$ lattice
with $N_{MC}=10^5$. 
    The true values of all coupling constants 
  were generated from a uniform distribution
  in the range $[ -0.5,0.5]$.
}
  \label{good assumption}
\end{center}
\end{figure}

Fig.~\ref{good assumption}
shows  that CFM
accurately reproduces the 
values of the coupling constants for
all types of interactions.
Although it is not obvious from this plot,
the errors 
in the four-spin coupling constants
are smaller than those for the two-spin couplings,
which are smaller than those for the local magnetic fields.

\begin{figure}[ht]
\begin{center}
 \includegraphics[width=\columnwidth]{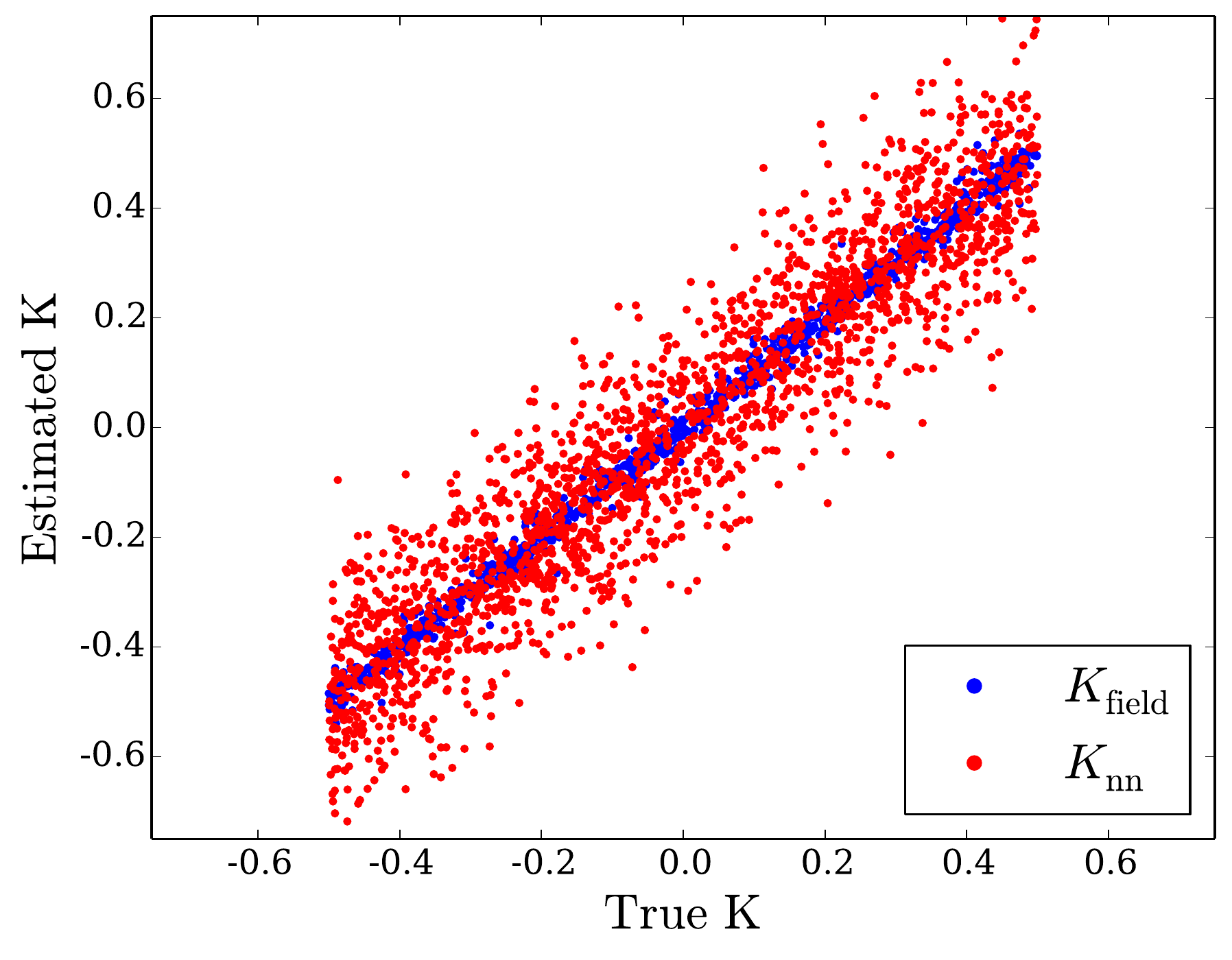}
  \caption{
  Plot of (badly) estimated  coupling constants 
  vs. the true values
  for the same MC simulation used in 
  Fig.~\ref{good assumption}.
  While the estimates of the coupling constants 
  only included nearest-neighbor interactions,
  the Hamiltonian of the MC simulation
 also  included  four-spin interactions.
Omitting the four-spin couplings
severely distorted the estimated values of 
$K_{field}$ and $K_{nn}$.}
  \label{bad assumption}
\end{center}
\end{figure}

Next,
we tried fitting our data with local magnetic fields and 
pairwise interactions,
but omitting the four-spin couplings.
Convergence was rapid,
and 
the local magnetizations and the 
two-spin correlation functions 
were fit to better than $10^{-10}$.
However,
as can be seen in 
Fig.~\ref{bad assumption},
the fitted values of the coupling constants 
deviated substantially from the true values.

\begin{figure}[t]
\begin{center}
 \includegraphics[width=\columnwidth]{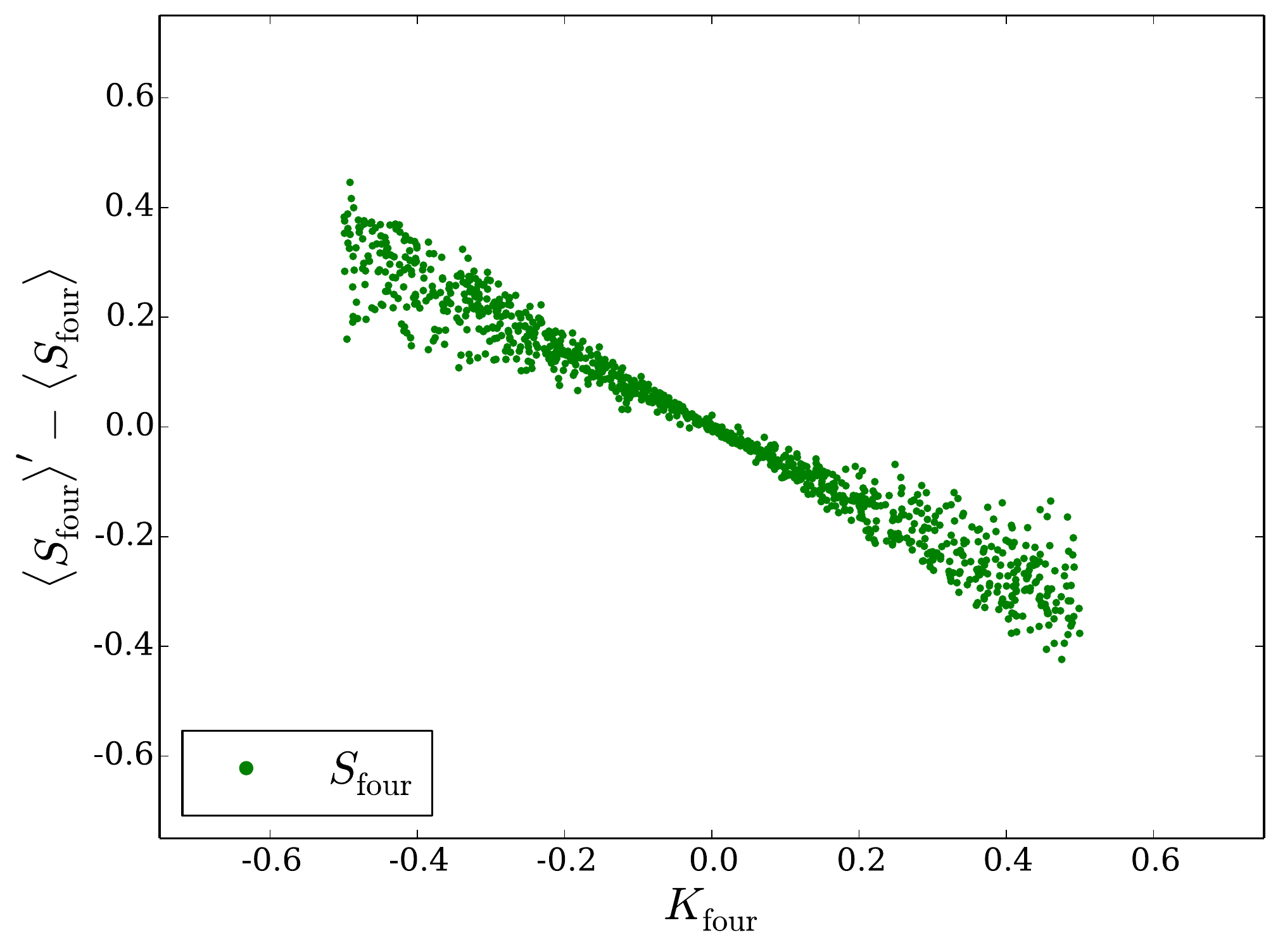}
  \caption{
  The two-spin coupling constants 
  and the local magnetic fields
  shown in Fig.~\ref{bad assumption}
  were used in a new MC simulation
  on the same
    $32 \times 32$ lattice 
  with $N_{MC}=2.5 \times 10^5$.
  This figure shows the 
  differences in the 
  four-spin correlation functions
  between the new simulation and the original data,
  plotted against the true 
  values of the
  four-spin coupling constants.
}
  \label{errors multi correlations}
\end{center}
\end{figure}

We  carried out a confirmation simulation,
using the local fields and two-spin couplings
(but no four-spin terms)
shown in Fig.~\ref{bad assumption}.
As expected,
we found good agreement 
with the 
local magnetizations and two-spin correlation functions 
from the full Hamiltonian (with four-spin terms).
However,
we found poor agreement for the 
four-spin correlations,
as shown in 
Fig.~\ref{errors multi correlations}.
  Although the two-spin correlation functions
  and the local magnetizations agree 
  to within the expected errors,
 there are significant
differences in the four-spin correlations.
For small values of the true four-spin coupling constants,
the deviations are nearly linear,
as expected.
However, 
the linear approximation 
becomes worse as the magnitude of the true coupling constants
increase.
These systematic deviations demonstrate 
the existence of multi-spin couplings neglected in the initial assumptions.

\FloatBarrier

\section{Conclusions}

We have shown that 
multi-spin coupling constants 
 can be accurately obtained 
in the inverse Ising problem
with the CFM approach\cite{RHS_MCRG_Ks,RHS_MCRG_Ks_d=2,RHS_MCRG_Ks_d=3}.
We have 
introduced and 
demonstrated 
confirmation testing,
which uses
 a new MC simulation to confirm (or deny)
whether 
a given set of effective coupling constants
provides
 a faithful representation 
of real data.

\makeatletter
\renewcommand\@biblabel[1]{#1. }
\makeatother

\bibliography{Citations_1}

%merlin.mbs 2010-03-15 4.21a (PWD, AO, DPC)
\begin{thebibliography}{14}%
\makeatletter
\providecommand \@ifxundefined [1]{%
 \@ifx{#1\undefined}
}%
\providecommand \@ifnum [1]{%
 \ifnum #1\expandafter \@firstoftwo
 \else \expandafter \@secondoftwo
 \fi
}%
\providecommand \@ifx [1]{%
 \ifx #1\expandafter \@firstoftwo
 \else \expandafter \@secondoftwo
 \fi
}%
\providecommand \natexlab [1]{#1}%
\providecommand \enquote  [1]{``#1''}%
\providecommand \bibnamefont  [1]{#1}%
\providecommand \bibfnamefont [1]{#1}%
\providecommand \citenamefont [1]{#1}%
\providecommand \href@noop [0]{\@secondoftwo}%
\providecommand \href [0]{\begingroup \@sanitize@url \@href}%
\providecommand \@href[1]{\@@startlink{#1}\@@href}%
\providecommand \@@href[1]{\endgroup#1\@@endlink}%
\providecommand \@sanitize@url [0]{\catcode `\\12\catcode `\$12\catcode
  `\&12\catcode `\#12\catcode `\^12\catcode `\_12\catcode `\%12\relax}%
\providecommand \@@startlink[1]{}%
\providecommand \@@endlink[0]{}%
\providecommand \url  [0]{\begingroup\@sanitize@url \@url }%
\providecommand \@url [1]{\endgroup\@href {#1}{\urlprefix }}%
\providecommand \urlprefix  [0]{URL }%
\providecommand \Eprint [0]{\href }%
\@ifxundefined \urlstyle {%
  \providecommand \doi  [0]{\begingroup \@sanitize@url \@doi}%
  \providecommand \@doi [1]{\endgroup \@@startlink {\doibase
  #1}doi:\discretionary {}{}{}#1\@@endlink }%
}{%
  \providecommand \doi  [0]{doi:\discretionary{}{}{}\begingroup
  \urlstyle{rm}\Url }%
}%
\providecommand \doibase [0]{http://dx.doi.org/}%
\providecommand \Doi [0]{\begingroup \@sanitize@url \@Doi }%
\providecommand \@Doi  [1]{\endgroup\@@startlink{\doibase#1}\@@Doi}%
\providecommand \@@Doi [1]{#1\@@endlink}%
\providecommand \selectlanguage [0]{\@gobble}%
\providecommand \bibinfo  [0]{\@secondoftwo}%
\providecommand \bibfield  [0]{\@secondoftwo}%
\providecommand \translation [1]{[#1]}%
\providecommand \BibitemOpen [0]{}%
\providecommand \bibitemStop [0]{}%
\providecommand \bibitemNoStop [0]{.\EOS\space}%
\providecommand \EOS [0]{\spacefactor3000\relax}%
\providecommand \BibitemShut  [1]{\csname bibitem#1\endcsname}%
%</preamble>
\bibitem [{\citenamefont {Schneidman}\ \emph {et~al.}(2008)\citenamefont
  {Schneidman}, \citenamefont {II}, \citenamefont {Segev},\ and\ \citenamefont
  {Bialek}}]{Schneidman}%
  \BibitemOpen
  \bibfield  {author} {\bibinfo {author} {\bibfnamefont {E.}~\bibnamefont
  {Schneidman}}, \bibinfo {author} {\bibfnamefont {M.~J.~Berry}\ \bibnamefont
  {II}}, \bibinfo {author} {\bibfnamefont {R.}~\bibnamefont {Segev}}, \ and\
  \bibinfo {author} {\bibfnamefont {W.}~\bibnamefont {Bialek}},\ }\bibfield
  {title} {\enquote {\bibinfo {title} {Weak pairwise correlations imply
  strongly correlated network states in a neural population},}\ }\href@noop {}
  {\bibfield  {journal} {\bibinfo  {journal} {Nature},\ }\textbf {\bibinfo
  {volume} {440}},\ \bibinfo {pages} {1007--1012} (\bibinfo {year}
  {2008})}\BibitemShut {NoStop}%
\bibitem [{\citenamefont {Lezon}\ \emph {et~al.}(2006)\citenamefont {Lezon},
  \citenamefont {Banavar}, \citenamefont {Cieplak}, \citenamefont {Maritan},\
  and\ \citenamefont {Fedoroff}}]{Lezon_2006}%
  \BibitemOpen
  \bibfield  {author} {\bibinfo {author} {\bibfnamefont {T.~R.}\ \bibnamefont
  {Lezon}}, \bibinfo {author} {\bibfnamefont {J.~R.}\ \bibnamefont {Banavar}},
  \bibinfo {author} {\bibfnamefont {M.}~\bibnamefont {Cieplak}}, \bibinfo
  {author} {\bibfnamefont {A.}~\bibnamefont {Maritan}}, \ and\ \bibinfo
  {author} {\bibfnamefont {N.~V.}\ \bibnamefont {Fedoroff}},\ }\bibfield
  {title} {\enquote {\bibinfo {title} {Using the principle of entropy
  maximization to infer genetic interaction networks from gene expression
  patterns},}\ }\href@noop {} {\bibfield  {journal} {\bibinfo  {journal} {Proc.
  Natl. Acad. Sci. U.S.A.},\ }\textbf {\bibinfo {volume} {103}},\ \bibinfo
  {pages} {19033--19038} (\bibinfo {year} {2006})}\BibitemShut {NoStop}%
\bibitem [{\citenamefont {Cocco}\ \emph {et~al.}(2009)\citenamefont {Cocco},
  \citenamefont {Leibler},\ and\ \citenamefont {Monasson}}]{Cocco_2009}%
  \BibitemOpen
  \bibfield  {author} {\bibinfo {author} {\bibfnamefont {S.}~\bibnamefont
  {Cocco}}, \bibinfo {author} {\bibfnamefont {S.}~\bibnamefont {Leibler}}, \
  and\ \bibinfo {author} {\bibfnamefont {R.}~\bibnamefont {Monasson}},\
  }\bibfield  {title} {\enquote {\bibinfo {title} {Neuronal couplings between
  retinal ganglion cells inferred by efficient inverse statistical physics
  methods,},}\ }\href@noop {} {\bibfield  {journal} {\bibinfo  {journal} {Proc.
  Natl. Acad. Sci. U.S.A.},\ }\textbf {\bibinfo {volume} {106}},\ \bibinfo
  {pages} {14058--14062} (\bibinfo {year} {2009})}\BibitemShut {NoStop}%
\bibitem [{\citenamefont {Tkacik}\ \emph {et~al.}(2006)\citenamefont {Tkacik},
  \citenamefont {Schneidman}, \citenamefont {II},\ and\ \citenamefont
  {Bialek}}]{Tkacik_Schneidman_Berry_Bialek_2006}%
  \BibitemOpen
  \bibfield  {author} {\bibinfo {author} {\bibfnamefont {G.}~\bibnamefont
  {Tkacik}}, \bibinfo {author} {\bibfnamefont {E.}~\bibnamefont {Schneidman}},
  \bibinfo {author} {\bibfnamefont {M.~J.~Berry}\ \bibnamefont {II}}, \ and\
  \bibinfo {author} {\bibfnamefont {W.}~\bibnamefont {Bialek}},\ }\href@noop {}
  {\enquote {\bibinfo {title} {Ising models for a network of real neurons},}\ }
  (\bibinfo {year} {2006}),\ \bibinfo {note} {arXiv:0912.5409}\BibitemShut
  {NoStop}%
\bibitem [{\citenamefont {Tkacik}\ \emph {et~al.}(2009)\citenamefont {Tkacik},
  \citenamefont {Schneidman}, \citenamefont {II},\ and\ \citenamefont
  {Bialek}}]{Tkacik_Schneidman_Berry_Bialek_2009}%
  \BibitemOpen
  \bibfield  {author} {\bibinfo {author} {\bibfnamefont {G.}~\bibnamefont
  {Tkacik}}, \bibinfo {author} {\bibfnamefont {E.}~\bibnamefont {Schneidman}},
  \bibinfo {author} {\bibfnamefont {M.~J.~Berry}\ \bibnamefont {II}}, \ and\
  \bibinfo {author} {\bibfnamefont {W.}~\bibnamefont {Bialek}},\ }\href@noop {}
  {\enquote {\bibinfo {title} {Spin glass models for a network of real
  neurons},}\ } (\bibinfo {year} {2009}),\ \bibinfo {note}
  {arXiv:0912.5409}\BibitemShut {NoStop}%
\bibitem [{\citenamefont {Weigt}\ \emph {et~al.}(2008)\citenamefont {Weigt},
  \citenamefont {White}, \citenamefont {Szurmant}, \citenamefont {Hoch},\ and\
  \citenamefont {Hwa}}]{Weigt_2008}%
  \BibitemOpen
  \bibfield  {author} {\bibinfo {author} {\bibfnamefont {M.}~\bibnamefont
  {Weigt}}, \bibinfo {author} {\bibfnamefont {R.~A.}\ \bibnamefont {White}},
  \bibinfo {author} {\bibfnamefont {H.}~\bibnamefont {Szurmant}}, \bibinfo
  {author} {\bibfnamefont {J.~A.}\ \bibnamefont {Hoch}}, \ and\ \bibinfo
  {author} {\bibfnamefont {T.}~\bibnamefont {Hwa}},\ }\bibfield  {title}
  {\enquote {\bibinfo {title} {Identification of direct residue contacts in
  proteinÐprotein interaction by message passing},}\ }\href@noop {} {\bibfield
  {journal} {\bibinfo  {journal} {Proc. Natl. Acad. Sci. U.S.A.},\ }\textbf
  {\bibinfo {volume} {106}},\ \bibinfo {pages} {67--72} (\bibinfo {year}
  {2008})}\BibitemShut {NoStop}%
\bibitem [{\citenamefont {Schug}\ \emph {et~al.}(2009)\citenamefont {Schug},
  \citenamefont {Weigt}, \citenamefont {Onuchic}, \citenamefont {Hwa},\ and\
  \citenamefont {Szurmant}}]{Schug_2009}%
  \BibitemOpen
  \bibfield  {author} {\bibinfo {author} {\bibfnamefont {A.}~\bibnamefont
  {Schug}}, \bibinfo {author} {\bibfnamefont {M.}~\bibnamefont {Weigt}},
  \bibinfo {author} {\bibfnamefont {J.~N.}\ \bibnamefont {Onuchic}}, \bibinfo
  {author} {\bibfnamefont {T.}~\bibnamefont {Hwa}}, \ and\ \bibinfo {author}
  {\bibfnamefont {H.}~\bibnamefont {Szurmant}},\ }\bibfield  {title} {\enquote
  {\bibinfo {title} {High-resolution protein complexes from integrating genomic
  information with molecular simulation},}\ }\href@noop {} {\bibfield
  {journal} {\bibinfo  {journal} {Proc. Natl. Acad. Sci. U.S.A.},\ }\textbf
  {\bibinfo {volume} {106}},\ \bibinfo {pages} {22124--22129} (\bibinfo {year}
  {2009})}\BibitemShut {NoStop}%
\bibitem [{\citenamefont {Morcos}\ \emph {et~al.}(2011)\citenamefont {Morcos},
  \citenamefont {Pagnani}, \citenamefont {Lunt}, \citenamefont {Bertolino},
  \citenamefont {Marks}, \citenamefont {Sander}, \citenamefont {Zecchina},
  \citenamefont {Onuchic}, \citenamefont {Hwa},\ and\ \citenamefont
  {Weigt}}]{Morcos_2011}%
  \BibitemOpen
  \bibfield  {author} {\bibinfo {author} {\bibfnamefont {F.}~\bibnamefont
  {Morcos}}, \bibinfo {author} {\bibfnamefont {A.}~\bibnamefont {Pagnani}},
  \bibinfo {author} {\bibfnamefont {B.}~\bibnamefont {Lunt}}, \bibinfo {author}
  {\bibfnamefont {A.}~\bibnamefont {Bertolino}}, \bibinfo {author}
  {\bibfnamefont {D.~S.}\ \bibnamefont {Marks}}, \bibinfo {author}
  {\bibfnamefont {C.}~\bibnamefont {Sander}}, \bibinfo {author} {\bibfnamefont
  {R.}~\bibnamefont {Zecchina}}, \bibinfo {author} {\bibfnamefont {J.~N.}\
  \bibnamefont {Onuchic}}, \bibinfo {author} {\bibfnamefont {T.}~\bibnamefont
  {Hwa}}, \ and\ \bibinfo {author} {\bibfnamefont {M.}~\bibnamefont {Weigt}},\
  }\bibfield  {title} {\enquote {\bibinfo {title} {Direct-coupling analysis of
  residue coevolution captures native contacts across many protein families},}\
  }\href@noop {} {\bibfield  {journal} {\bibinfo  {journal} {Proc. Natl. Acad.
  Sci. U.S.A.},\ }\textbf {\bibinfo {volume} {108}},\ \bibinfo {pages}
  {E1293--E1301} (\bibinfo {year} {2011})}\BibitemShut {NoStop}%
\bibitem [{\citenamefont {Aurell}\ and\ \citenamefont
  {Ekeberg}(2012)}]{Aurell_Ekeberg}%
  \BibitemOpen
  \bibfield  {author} {\bibinfo {author} {\bibfnamefont {E.}~\bibnamefont
  {Aurell}}\ and\ \bibinfo {author} {\bibfnamefont {M.}~\bibnamefont
  {Ekeberg}},\ }\bibfield  {title} {\enquote {\bibinfo {title} {Inverse ising
  inference using all the data},}\ }\href@noop {} {\bibfield  {journal}
  {\bibinfo  {journal} {Phys. Rev. Lett.},\ }\textbf {\bibinfo {volume}
  {108}},\ \bibinfo {pages} {090201} (\bibinfo {year} {2012})}\BibitemShut
  {NoStop}%
\bibitem [{\citenamefont {Swendsen}(1984){\natexlab{a}}}]{RHS_MCRG_Ks}%
  \BibitemOpen
  \bibfield  {author} {\bibinfo {author} {\bibfnamefont {R.~H.}\ \bibnamefont
  {Swendsen}},\ }\bibfield  {title} {\enquote {\bibinfo {title} {{Monte}
  {Carlo} calculation of renormalized coupling parameters},}\ }\href@noop {}
  {\bibfield  {journal} {\bibinfo  {journal} {Phys, Rev. Letters},\ }\textbf
  {\bibinfo {volume} {52}},\ \bibinfo {pages} {1165} (\bibinfo {year}
  {1984}{\natexlab{a}})}\BibitemShut {NoStop}%
\bibitem [{\citenamefont {Swendsen}(1984){\natexlab{b}}}]{RHS_MCRG_Ks_d=2}%
  \BibitemOpen
  \bibfield  {author} {\bibinfo {author} {\bibfnamefont {R.~H.}\ \bibnamefont
  {Swendsen}},\ }\bibfield  {title} {\enquote {\bibinfo {title} {{Monte}
  {Carlo} calculation of renormalized coupling parameters: {I}. d=2 {I}sing
  model},}\ }\href@noop {} {\bibfield  {journal} {\bibinfo  {journal} {Phys.
  Rev. B},\ }\textbf {\bibinfo {volume} {30}},\ \bibinfo {pages} {3866}
  (\bibinfo {year} {1984}{\natexlab{b}})}\BibitemShut {NoStop}%
\bibitem [{\citenamefont {Swendsen}(1984){\natexlab{c}}}]{RHS_MCRG_Ks_d=3}%
  \BibitemOpen
  \bibfield  {author} {\bibinfo {author} {\bibfnamefont {R.~H.}\ \bibnamefont
  {Swendsen}},\ }\bibfield  {title} {\enquote {\bibinfo {title} {{M}onte
  {C}arlo calculation of renormalized coupling parameters: {II}. d=3 {I}sing
  model},}\ }\href@noop {} {\bibfield  {journal} {\bibinfo  {journal} {Phys.
  Rev. B},\ }\textbf {\bibinfo {volume} {30}},\ \bibinfo {pages} {3875}
  (\bibinfo {year} {1984}{\natexlab{c}})}\BibitemShut {NoStop}%
\bibitem [{\citenamefont {Callen}(1985)}]{Callen}%
  \BibitemOpen
  \bibfield  {author} {\bibinfo {author} {\bibfnamefont {H.~B.}\ \bibnamefont
  {Callen}},\ }\href@noop {} {\emph {\bibinfo {title} {Thermodynamics and an
  Introduction to Thermostatistics}}},\ \bibinfo {edition} {2nd}\ ed.\
  (\bibinfo  {publisher} {Wiley},\ \bibinfo {address} {New York},\ \bibinfo
  {year} {1985})\BibitemShut {NoStop}%
\bibitem [{\citenamefont {Sherrington}\ and\ \citenamefont
  {Kirkpatrick}(1975)}]{SK}%
  \BibitemOpen
  \bibfield  {author} {\bibinfo {author} {\bibfnamefont {D.}~\bibnamefont
  {Sherrington}}\ and\ \bibinfo {author} {\bibfnamefont {S.}~\bibnamefont
  {Kirkpatrick}},\ }\bibfield  {title} {\enquote {\bibinfo {title} {Solvable
  model of a spin-glass},}\ }\href@noop {} {\bibfield  {journal} {\bibinfo
  {journal} {Phys. Rev. Lett.},\ }\textbf {\bibinfo {volume} {35}},\ \bibinfo
  {pages} {1792--1796} (\bibinfo {year} {1975})}\BibitemShut {NoStop}%
\end{thebibliography}%

\end{document}